# The Trace Formula of the Spinoriel Amplitude

Mustapha MEKHFI

International Center for Theoretical Physics, Trieste, ITALY

We re express the fermion's probability amplitude as a trace over spinor indices which reformulation surprisingly does not exist in literature. This reformulation puts the probability amplitude and the probability (squared amplitude) of a given process on equal footing at the algebraic computation level and this is our principal motivation to write the paper. We test the power of the trace formula in three applications: Calculation of the charge-current of fermions by using symbolic programs which current so far was only computable by hand, analytic computation of the quark dipole magnetic moment, rendered less cumbersome, and finally Fiertz rearrangement identities now made more transparent.

# 1 Introduction

The probability amplitude of a process involving fermions of spin one half are generally written as

$$\bar{u}(p',s')...f.....u(p,s) \tag{1}$$

$u(p,s)$ is the part of the wave function describing the spin of a particle of energy momentum $p$ and spin $s$ and ellipses indicate other Dirac spinors. Cross sections and lifetimes are two basic observables in particle and nuclear physics. These observables are built out of the probability for the related process to occur, in addition to kinematical variables. The probability for a given process to occur is the squared modulus of the probability amplitude. It is usually expressed as a trace over spinor indices by use of projectors

$$Tr(\bar{f}\frac{\slashed{p}'+m}{2m}\frac{1+\gamma_5\slashed{s}'}{2}\cdots f\cdots\frac{\slashed{p}+m}{2m}\frac{1+\gamma_5\slashed{s}}{2}) \tag{2}$$

The trace form of the probability is compact and Lorentz covariant, and in addition it offers the possibility to handle expressions involving several $\gamma$- matrices (several loops) using machine facilities. Several symbolic programs are made available for such symbolic computations. In the next section we will show that the probability amplitudes can themselves be re-expressed as traces as well; and hence will benefit from the same computational facilities as probabilities.

## 2 The probability amplitude as a trace

Cross sections and lifetimes are not the only observables of interest. Other observables such as form factors, and particularly magnetic or electric dipole moments, etc , are also of interest. These are however probability amplitudes rather than probabilities. We propose in this paper to rewrite the probability amplitude as a trace over spinor indices, as in(2). In this way the probability amplitude gets a compact form in terms of "generalized" projectors, thus

becoming suitable for analysis and ready for symbolic computations. To this end we rewrite (1) as

$$\begin{aligned}&\bar{u}_{\alpha'}(k',s')\cdots f_{\alpha'\alpha}\cdots u_{\alpha}(k,s)\\&=\cdots f_{\alpha'\alpha}\cdots u_{\alpha}(k,s)\bar{u}_{\alpha'}(k',s')\\&=Tr(\cdots f\rho)\\&\rho_{\alpha\alpha'}(k,s,k',s')=u_{\alpha}(k,s)\bar{u}_{\alpha'}(k',s')\end{aligned} \quad (3)$$

The trace involves the 4 by 4 matrix $\rho_{\alpha\alpha'}(k,s,k',s')$ ( generalized spin density matrix ) which reduces to the usual product of projectors

$$(\frac{\not{k}+m}{2m})(\frac{1+\gamma^5\not{s}}{2}) \quad (4)$$

When $k = k'$ and $s = s'$, and being a matrix in the space of $\gamma$- matrices, will be re expressed in terms of $\gamma$- matrices. Two methods will be proposed to compute the spin density matrix. The first one is a generalization of the method[1] used to work out the product of projectors(4). This method uses extensively the properties of the spin density matrix which may be inferred from its very definition

$$\begin{aligned}&\rho_{\alpha\alpha'}(k,s,k',s')=u_{\alpha}(k,s)\bar{u}_{\alpha'}(k',s')\\&(\not{k}-m)\rho(k,s,k',s')=0=\rho(k,s,k',s')(\not{k}'-m)\\&\rho(k,s,k',s')u(k',s')=u(k,s)\\&\bar{u}(k,s)\rho(k,s,k',s')=\bar{u}(k',s')\end{aligned} \quad (5)$$

The second method is the method[2] used to work out the product of projectors(4) by relating the matrix $\rho$ directly to its form within the rest frame (easy to compute). Then to get $\rho$, we perform a Lorentz boost to the initial frame. Both methods lead to a density of the form

$$\rho(k,s,k',s')=(\frac{\not{k}+m}{2m})(\frac{1+\gamma^5\not{s}}{2})\Re(\frac{\not{k}'+m}{2m})(\frac{1+\gamma^5\not{s}'}{2}) \quad (6)$$

The matrix $\Re$ obtained after factorizing the projectors from the spin density has the explicit form

$$\Re = \cosh(\frac{\omega}{2})\cosh(\frac{\omega'}{2})$$
$$-\sinh(\frac{\omega}{2})\cosh(\frac{\omega'}{2})\gamma_0\frac{\vec{\gamma}.\vec{k}}{|\vec{k}|} + \sinh(\frac{\omega'}{2})\cosh(\frac{\omega}{2})\gamma_0\frac{\vec{\gamma}.\vec{k}'}{|\vec{k}'|} \quad (7)$$
$$+\sinh(\frac{\omega}{2})\sinh(\frac{\omega'}{2})\frac{\vec{\gamma}.\vec{k}}{|\vec{k}|}\frac{\vec{\gamma}.\vec{k}'}{|\vec{k}'|}$$

With $\omega = -\tanh^{-1}(\frac{|\vec{k}|}{k^0})$ and idem for $\omega'$. Full details of the computation of the spin density $\rho$ will be given in appendix A.

## 3 Applications

**3.1 Symbolic computations of the charge-current**

One of the difficulties in computing a Feynman diagram is trace calculations. So the calculations of Dirac $\gamma$- matrices were one of first task of computer algebra systems. Many powerful algorithms[3] are now available for calculating them. These algorithms are implemented in various user friendly software packages. Among the most popular is FeynCalc[4]. FeynCalc is a Mathematica[5] package for algebraic calculations in high energy physics. It provides tools for Lorentz structure manipulations, Dirac algebra manipulations, color factor calculations, Feynman rule derivation, Feynman loop integral calculations etc. There is also Maple[6], Reduce[7], Maxima[8] etc. All these powerfull computation tools are now made able (by the trace formula for amplitudes) to compute spin amplitudes phrased as traces. The form of the amplitude adequate for symbolic computation is

$$Tr(\rho f) \quad (8)$$

With the generalized spin matrix density $\rho$, a kinematical quantity entirely fixed in(6). To illustrate the approach we propose to compute the fermion charge- current, first by hand as usual, then by the use of machine facilities we re compute the current with the help of the trace formula. To compute the current $\bar{u}(\vec{k}',s')\vec{\gamma}u(\vec{k},s)$ by hand and to be simple we work in the limit $k'_0 = k_0$ (transfer of space-like type). The result we get reads

$$\bar{u}(\vec{k}',s')\vec{\gamma}u(\vec{k},s) = (\frac{1+\vec{s}.\vec{s}'}{4m})(\vec{k}+\vec{k}') + i(\vec{k}-\vec{k}') \times \frac{(\vec{s}+\vec{s}'+i\vec{s}\times\vec{s}')}{4m} \qquad (9)$$

We may further simplify the expression as follows: we average over the initial spin, put the initial momentum along the z axis and look to the z component of the current. After these simplifications we get the simple result (by hand).

$$\frac{1}{2}\sum_s \bar{u}(\vec{k}',s')\gamma_3 u(\vec{k},s) = \frac{1}{2}\sum_s Tr(\rho\gamma_3) = \frac{(k_3+k_3') + i(-k'_1 s_2' + k'_2 s_1')}{4m} \qquad (10)$$

For the symbolic computation, we propose a simple program written in Maple. We use a special package[9] (Maple 7 and higher now Maple 13) for the calculation of traces of Dirac gamma matrices and get the result(10) (see Appendix B). It should be stressed that this is the first time one compute an amplitude algebraically.

### 3.2 Dipole magnetic moment of the quark: The convection current

The calculation of the dipole magnetic moment is an adequate place for another relatively interesting application of the trace formula. To illustrate it we just content ourselves with the computation of the convection current contribution to the dipole magnetic moment. In the momentum space this contribution has the following form[10]

$$\frac{Q}{2m}\int\left[\vec{\nabla}_q Tr\rho\right]_{\vec{q}=0} \times \vec{k} \, \frac{d^3k}{(2\pi)^3} \qquad (11)$$

$$\rho_{\alpha\alpha'}(k,s,k',s') = u_\alpha(k,s)\bar{u}_{\alpha'}(k's')$$

(we omit the normalization factors $\sqrt{\frac{m}{k_0}}$ for each spinor and write only $u$ instead of $\psi = \sqrt{\frac{m}{k_0}}u$ and denote $\vec{q} = \vec{k} - \vec{k}'$ ).To compute(11), we insert the expression of $\rho$ computed in(6), perform the differentiation with respect to $\vec{q}$ and set $\vec{q} = 0$. A look at the expression in(11) reveals that the dependence on the momentum transfer $\vec{q}$ is only within the matrix $\Re$. Hence, we need to differentiate $\Re$. This is physical as for $\Re=0$ ($\vec{k}=\vec{k}\,'$) there is no acceleration of the charge and hence no magnetism. The explicit calculation is as follows.

$$\vec{k} \times \vec{\nabla}_q \Re |_{\vec{q}=0} = -\frac{1}{|\vec{k}|} \sinh(\frac{\omega}{2}) \cosh(\frac{\omega}{2}) \vec{k} \times \gamma_0 \vec{\gamma}$$
$$-\frac{1}{2|\vec{k}\,'||\vec{k}|} \sinh^2(\frac{\omega}{2}) \vec{k} \times (\vec{\gamma}\vec{\gamma}.\vec{k} - \vec{\gamma}.\vec{k}\vec{\gamma})$$
$$= -\frac{1}{|\vec{k}|} \sinh(\frac{\omega}{2}) \cosh(\frac{\omega}{2}) \vec{k} \times \gamma_0 \vec{\gamma}$$
$$+\frac{k_0}{|\vec{k}|^2} \sinh^2(\frac{\omega}{2}) \vec{k} \times \gamma_0 \vec{\gamma} \quad (12)$$
$$= \left[ -\frac{1}{|\vec{k}|} \sinh(\frac{\omega}{2}) \cosh(\frac{\omega}{2}) + \frac{k_0}{|\vec{k}\,'||\vec{k}|} \sinh^2(\frac{\omega}{2}) \right] \vec{k} \times \gamma_0 \vec{\gamma}$$
$$= \frac{1}{2} \frac{1}{k_0 + m} \vec{k} \times \gamma_0 \vec{\gamma}$$

Inserting the result(12) back into(11) we get the convection current contribution to the dipole magnetic moment of the quark

$$-\int \vec{k} \times \left[ \vec{\nabla} Tr\rho \right]_{\vec{q}=0} \frac{d^3k}{(2\pi)^3}$$
$$= -\frac{1}{2} \frac{1}{(k_0+m)} \int (\vec{k} \times Tr \left[ (\frac{\not{k}+m}{2m})(\frac{1+\gamma_5 \not{s}}{2})(\gamma_0 \vec{\gamma}) \right]) \frac{d^3k}{(2\pi)^3}$$
$$= \frac{1}{8m} \frac{1}{(k_0+m)} \int (\vec{k} \times Tr \gamma_5 \not{k} \not{s} \gamma_0 \vec{\gamma}) \frac{d^3k}{(2\pi)^3} \quad (13)$$
$$= -\frac{1}{2m} \frac{1}{(k_0+m)} \int |\vec{k}|^2 \vec{s}_\perp \frac{d^3k}{(2\pi)^3}$$
$$= -\frac{x}{(x+1)} \int \frac{|\vec{k}|^2}{2m^2} \vec{s}_\perp \frac{d^3k}{(2\pi)^3}$$

In the equation above, we approximate the factor $k_0$ by the average value of the relativistic quark energy inside the nucleon and keep the notation $k_0$ to designate the average value, while the parameter $x$ is the ratio $x = \frac{m}{k_0}$. At this level, we can show that the integral in the last line of(13) is the combination of the spin operators $\vec{S}$ and $\vec{\delta}$ commonly called longitudinal and transverse spin respectively. (This naming is accurate only at high energy,

where $\frac{m}{k_0} \ll 1$ as is evidenced from(14) (we skip the details of the computation below to

lighten the paper)

$$\begin{aligned}
\vec{S} - \frac{\vec{\delta}}{2x} &= -\frac{1}{2}\int \frac{|k|^2}{m^2} \vec{s}_\perp(k) \frac{d^3k}{(2\pi)^3} \\
\vec{S} &= \frac{1}{2}\int \bar{\psi}\vec{\gamma}\gamma_5\psi \frac{d^3k}{(2\pi)^3} = \frac{1}{2}\int (\frac{k^0}{m}\zeta_\| \vec{n} + \vec{s}_\perp(k))\frac{d^3k}{(2\pi)^3} \\
\vec{\delta} &= \int \bar{\psi}\vec{\gamma}\gamma_5\gamma_0\psi \frac{d^3k}{(2\pi)^3} = \int (\zeta_\| \vec{n} + \frac{k_0}{m}\vec{s}_\perp)\frac{d^3k}{(2\pi)^3}
\end{aligned} \quad (14)$$

Putting together results from(13) ,(14) and re establishing the normalization factors into(13) ,

we get the convection current part of the quark dipole magnetic moment [7]

$$\begin{aligned}
&\frac{x\mu}{(x+1)}(\vec{S} - \frac{\vec{\delta}}{2x}) \\
&\mu = \frac{Q}{2m}
\end{aligned} \quad (15)$$

To get a feeling of the power of the trace formulation one has to compare this rapid and clear derivation of the result(15) with the identical result obtained in references[11,12] after lengthy and cumbersome direct computations (based on the explicit form of the Dirac spinors ).

### 3.3 Fiertz rearrangement

The trace formula can be applied here and has the advantage of making Fiertz identities much

more transparent. We note however that only the reformulation of spinors in terms of traces of

Dirac gamma matrices and the subsequent properties of the latter ones are necessary to derive

the various identities, no explicit form of $\rho$ is needed. An example of a Fiertz identity is

(indices $i, j, k, l$ may indicate energy-momentum or any other quantum number)

$$\begin{aligned}
(\bar{u}_i u_j)(\bar{u}_k u_l) &= Tr(\rho_{jk}\rho_{li}) \\
&= \sum_{A,B} \rho_{jkA}\rho_{li}{}^B Tr(\gamma^A \gamma_B) = 4\sum_{A,B} \rho_{jkA}\rho_{li}{}^B \delta_B^A = 4\sum_A \rho_{jkA}\rho_{li}{}^A \\
&= \frac{1}{4}\sum_A Tr(\rho_{jk}\gamma_A)Tr(\rho_{li}\gamma^A) \\
&= \frac{1}{4}\sum_A (\bar{u}_k \gamma_A u_j)(\bar{u}_i \gamma^A u_l)
\end{aligned} \quad (16)$$

## 4  Conclusion

We reformulated the fermion probability amplitude as a trace over spinor indices. In this way, we put the probability amplitude and the probability (squared amplitude) of a given process on equal footing at the algebraic computation level. We illustrated our trace formula in three different areas. One is symbolic computation: The symbolic computation of the probability amplitude is rendered possible (most algebraic programs are conceived to compute a trace). The other is the use of the trace formula in computing the dipole magnetic moment: The dependence on the transfer momentum of the process is entirely in the matrix $\mathfrak{R}$, hence the ease in finding the result, and finally a minor application to Fiertz rearrangement: Fiertz identities when re-expressed as traces become more transparent.

## Appendix A )

## First method

The matrix $\rho_{\alpha\alpha'}(k,s,k',s') = u_\alpha(k,s)\bar{u}_{\alpha'}(k',s')$ obeys by definition the following equalities

$$(\slashed{k}-m)\rho(k,s,k',s') = 0 = \rho(k,s,k',s')(\slashed{k}'-m)$$
$$\rho(k,s,k',s')u(k',s') = u(k,s) \qquad (A18)$$
$$\bar{u}(k,s)\rho(k,s,k',s') = \bar{u}(k',s')$$

As a consequence of the first equation in (A18) $\rho(k,s,k',s')$ should have the form

$$\rho(k,s,k',s') = (\frac{\slashed{k}+m}{2m})\mathfrak{R}'(s,s')(\frac{\slashed{k}'+m}{2m}) \qquad (A.18)$$

At this point we note that the spin part of the projector $(\frac{\slashed{k}+m}{2m})$ is missing; because the spin is also involved here. We thus guess that $\mathfrak{R}'(s,s')$ should factorize as follows

$$\mathfrak{R}'(k,s,k',s') = (\frac{1+\gamma^5\slashed{s}}{2})\mathfrak{R}(k,k')(\frac{1+\gamma^5\slashed{s}'}{2}) \qquad (A.19)$$

Putting (A.19) and (A.18) together gives

$$\rho(k,s,k',s') = (\frac{\slashed{k}+m}{2m})(\frac{1+\gamma^5\slashed{s}}{2})\mathfrak{R}(\frac{\slashed{k}'+m}{2m})(\frac{1+\gamma^5\slashed{s}'}{2}) \qquad (A.20)$$

It remains to determine the matrix $\mathfrak{R}$. Using the second and the third equation in (A18) and the property of the projector: $(\frac{\slashed{k}+m}{2m})(\frac{1+\lambda\gamma^5\slashed{s}}{2})u(k,s) = u(k,s)$ for both $k$ and $k'$, we infer that the matrix $\mathfrak{R}$ should obey the relations

$$\mathfrak{R}(k,k')u(k',s') = u(k,s)$$
$$\bar{u}(k,s)\mathfrak{R}(k,k') = \bar{u}(k',s') \qquad (A.21)$$

$\mathfrak{R}$ is thus the operator responsible for the flip of the momentum and of the spin. The natural way for $\mathfrak{R}$ to operate the change in (A.21), is to boost the spinor $u(k',s')$ from $\vec{k}'$ down to the rest frame and then to boost it up again to the new momentum $\vec{k}$ to yield the spinor $u(k,s)$. $\mathfrak{R}$

should thus have the form $\Re(\vec{k},\vec{k}') = \Lambda(\vec{k})\Lambda^{-1}(\vec{k}')$, where $\Lambda(k) = \exp(-\frac{\omega}{2}\gamma_0 \frac{\vec{\gamma}.\vec{k}}{|\vec{k}|})$ is the boost operator (from the rest frame up to the momentum $\vec{k}^2$), with the Lorentz angle $\omega = -\tanh^{-1}(\frac{|\vec{k}|}{k^0})$. The explicit form of $\Re$ is thus

$$\Re = \exp(-\frac{\omega}{2}\gamma_0 \frac{\vec{\gamma}.\vec{k}}{|\vec{k}|}) \exp(\frac{\omega'}{2}\gamma_0 \frac{\vec{\gamma}.\vec{k}'}{|\vec{k}'|}) \tag{A.22}$$

The exponential has an argument with the property $\gamma_0 \frac{\vec{\gamma}.\vec{k}}{|\vec{k}|} \gamma_0 \frac{\vec{\gamma}.\vec{k}}{|\vec{k}|} = 1$. We then use this property to simplify the exponential

$$\exp(-\frac{\omega}{2}\gamma_0 \frac{\vec{\gamma}.\vec{k}}{|\vec{k}|}) = \cosh(\frac{\omega}{2}) - \sinh(\frac{\omega}{2})\gamma_0 \frac{\vec{\gamma}.\vec{k}}{|\vec{k}|} \tag{A.23}$$

and hence, the expression of $\Re$ gets simplified too

$$\begin{aligned}\Re &= \cosh(\frac{\omega}{2})\cosh(\frac{\omega'}{2}) \\ &-\sinh(\frac{\omega}{2})\cosh(\frac{\omega'}{2})\gamma_0 \frac{\vec{\gamma}.\vec{k}}{|\vec{k}|} + \sinh(\frac{\omega'}{2})\cosh(\frac{\omega}{2})\gamma_0 \frac{\vec{\gamma}.\vec{k}'}{|\vec{k}'|} \\ &+\sinh(\frac{\omega}{2})\sinh(\frac{\omega'}{2}) \frac{\vec{\gamma}.\vec{k}}{|\vec{k}|} \frac{\vec{\gamma}.\vec{k}'}{|\vec{k}'|}\end{aligned} \tag{A.24}$$

### Second method

This is another variant of the last computation; it is more direct, rigorous, but relatively lengthy. The idea is to relate the matrix $\rho$ (we raise all indices to simplify notation) directly to its form within the rest frame, where this form is easy to compute. Then to get $\rho$, we perform a Lorentz boost to the initial frame. So let us rewrite the Dirac spinor and it complex conjugate as:

$$u(k,s) = \exp(-\frac{\omega}{2}\gamma_0 \frac{\vec{\gamma}.\vec{k}}{|\vec{k}|})\begin{pmatrix} \chi \\ 0 \end{pmatrix}$$

$$\bar{u}(k,s) = \begin{pmatrix} \chi^\dagger, & 0 \end{pmatrix} \exp(-\frac{\omega}{2}\gamma_0 \frac{\vec{\gamma}.\vec{k}}{|\vec{k}|})\gamma_0$$

(A.25)

$\rho$ can now be re-written in terms of its expression in the rest frame $\chi\tilde{\chi}^\dagger$.

$$\rho = \exp(-\frac{\omega}{2}\gamma_0 \frac{\vec{\gamma}.\vec{k}}{|\vec{k}|})\begin{pmatrix} \chi \\ 0 \end{pmatrix}\begin{pmatrix} \tilde{\chi}^\dagger, & 0 \end{pmatrix}\exp(\frac{\omega'}{2}\gamma_0 \frac{\vec{\gamma}.\vec{k}'}{|\vec{k}'|})$$

$$= \exp(-\frac{\omega}{2}\gamma_0 \frac{\vec{\gamma}.\vec{k}}{|\vec{k}|})\begin{pmatrix} \chi\tilde{\chi}^\dagger & 0 \\ 0 & 0 \end{pmatrix}\exp(\frac{\omega'}{2}\gamma_0 \frac{\vec{\gamma}.\vec{k}'}{|\vec{k}'|})$$

(A.26)

We then work out the expression of $\chi\tilde{\chi}^\dagger$

$$\chi(\vec{\zeta})\tilde{\chi}^\dagger(\vec{\zeta}\,') = \frac{1+\vec{\zeta}.\vec{\zeta}\,'}{4} + \frac{(\vec{\zeta}+\vec{\zeta}\,'+i\vec{\zeta}\times\vec{\zeta}\,')}{4}.\vec{\sigma}$$

$$= (\frac{1+\vec{\zeta}.\vec{\sigma}}{2})(\frac{1+\vec{\zeta}\,'.\vec{\sigma}}{2})$$

(A.27)

In the above equation, we have re established the arguments of the product $\chi\tilde{\chi}^\dagger$.

Inserting (A.27) into (A.26) we get

$$\rho = \exp(-\frac{\omega}{2}\gamma_0 \frac{\vec{\gamma}.\vec{k}}{|\vec{k}|})\begin{pmatrix} (\frac{1+\vec{\zeta}.\vec{\sigma}}{2})(\frac{1+\vec{\zeta}\,'.\vec{\sigma}}{2}) & 0 \\ 0 & 0 \end{pmatrix}\exp(\frac{\omega'}{2}\gamma_0 \frac{\vec{\gamma}.\vec{k}'}{|\vec{k}'|})$$

$$= \exp(-\frac{\omega}{2}\gamma_0 \frac{\vec{\gamma}.\vec{k}}{|\vec{k}|})\begin{pmatrix} (\frac{1+\vec{\zeta}.\vec{\sigma}}{2}) & 0 \\ 0 & 0 \end{pmatrix}\begin{pmatrix} (\frac{1+\vec{\zeta}\,'.\vec{\sigma}}{2}) & 0 \\ 0 & 0 \end{pmatrix}\exp(\frac{\omega'}{2}\gamma_0 \frac{\vec{\gamma}.\vec{k}'}{|\vec{k}'|})$$

(A.28)

Now we insert the identity 1(written as the product of four exponentials) between the two matrices in (A.28), and get

$$\rho = \exp(-\frac{\omega}{2}\gamma_0 \frac{\vec{\gamma}.\vec{k}}{|\vec{k}|}) \begin{pmatrix} (\frac{1+\vec{\zeta}.\vec{\sigma}}{2}) & 0 \\ 0 & 0 \end{pmatrix} \exp(\frac{\omega}{2}\gamma_0 \frac{\vec{\gamma}.\vec{k}}{|\vec{k}|})$$

$$\exp(-\frac{\omega}{2}\gamma_0 \frac{\vec{\gamma}.\vec{k}}{|\vec{k}|}) \exp(\frac{\omega'}{2}\gamma_0 \frac{\vec{\gamma}.\vec{k}'}{|\vec{k}'|}) \qquad (A.29)$$

$$\exp(-\frac{\omega'}{2}\gamma_0 \frac{\vec{\gamma}.\vec{k}'}{|\vec{k}'|}) \begin{pmatrix} (\frac{1+\vec{\zeta}'.\vec{\sigma}}{2}) & 0 \\ 0 & 0 \end{pmatrix} \exp(\frac{\omega'}{2}\gamma_0 \frac{\vec{\gamma}.\vec{k}'}{|\vec{k}'|})$$

The second line in (A.29) is already the $\Re$ matrix. To get the desired result(6) it remains to show that the first and the second lines in (A.29) are the projectors (idem for $k'$)

$$\exp(-\frac{\omega}{2}\gamma_0 \frac{\vec{\gamma}.\vec{k}}{|\vec{k}|}) \begin{pmatrix} (\frac{1+\vec{\zeta}.\vec{\sigma}}{2}) & 0 \\ 0 & 0 \end{pmatrix} \exp(\frac{\omega}{2}\gamma_0 \frac{\vec{\gamma}.\vec{k}}{|\vec{k}|})$$
$$= (\frac{\slashed{k}+m}{2m})(\frac{1+\gamma^5 \slashed{s}}{2}) \qquad (A.30)$$

To this end we first compute the projector without spin (the projector on the positive energy state).

$$\exp(-\frac{\omega}{2}\gamma_0 \frac{\vec{\gamma}.\vec{k}}{|\vec{k}|}) \begin{pmatrix} 1 & 0 \\ 0 & 0 \end{pmatrix} \exp(\frac{\omega}{2}\gamma_0 \frac{\vec{\gamma}.\vec{k}}{|\vec{k}|})$$
$$= \exp(-\frac{\omega}{2}\gamma_0 \frac{\vec{\gamma}.\vec{k}}{|\vec{k}|})(\frac{1+\gamma_0}{2})\exp(\frac{\omega}{2}\gamma_0 \frac{\vec{\gamma}.\vec{k}}{|\vec{k}|})$$
$$= \frac{1}{2} + \frac{1}{2}\gamma_0 \exp(\omega \gamma_0 \frac{\vec{\gamma}.\vec{k}}{|\vec{k}|})$$
$$= \frac{1}{2} + \frac{1}{2}\gamma_0 (\cosh(\omega) + \sinh(\omega)\gamma_0 \frac{\vec{\gamma}.\vec{k}}{|\vec{k}|}) \qquad (A.31)$$
$$= \frac{1}{2} + \frac{1}{2}\gamma_0 (\frac{k_0}{m} - \frac{|k|}{m}\gamma_0 \frac{\vec{\gamma}.\vec{k}}{|\vec{k}|}) = \frac{1}{2} + \frac{1}{2m}(\gamma_0 k_0 - \vec{\gamma}.\vec{k})$$
$$= \frac{\slashed{k}+m}{2m}$$

Then we compute the spin part:

$$\exp(-\frac{\omega}{2}\gamma_0\frac{\vec{\gamma}.\vec{k}}{|\vec{k}|})\begin{pmatrix}\vec{\zeta}.\vec{\sigma} & 0 \\ 0 & 0\end{pmatrix}\exp(\frac{\omega}{2}\gamma_0\frac{\vec{\gamma}.\vec{k}}{|\vec{k}|})$$

$$= (\cosh(\omega) - \sinh(\omega)\gamma_0\frac{\vec{k}.\vec{\gamma}}{|\vec{k}|})(\frac{1+\gamma_0}{2})(\vec{\gamma}.\vec{\zeta})\gamma_5(\cosh(\omega) + \sinh(\omega)\gamma_0\frac{\vec{k}.\vec{\gamma}}{|\vec{k}|})$$

(A.32)

Let us move $\vec{\gamma}.\vec{\zeta}\gamma_5$ to the right in (A.32):

$$(\vec{\gamma}.\vec{\zeta})\gamma_5(\cosh(\omega) + \sinh(\omega)\gamma_0\frac{\vec{k}.\vec{\gamma}}{|\vec{k}|})$$

$$= (\cosh(\omega) + \sinh(\omega)\gamma_0\frac{\vec{k}.\vec{\gamma}}{|\vec{k}|})(\vec{\gamma}.\vec{\zeta})\gamma_5 + 2\sinh(\omega)\gamma_0\gamma_5\frac{(\vec{k}.\vec{\zeta})}{|\vec{k}|}$$

(A.33)

Inserting (A.33) back into (A.32) we get:

$$(\cosh(\omega) - \sinh(\omega)\gamma_0\frac{\vec{k}.\vec{\gamma}}{|\vec{k}|})(\frac{1+\gamma_0}{2})(\vec{\gamma}.\vec{\zeta})\gamma_5(\cosh(\omega) + \sinh(\omega)\gamma_0\frac{\vec{k}.\vec{\gamma}}{|\vec{k}|})$$

$$= (\cosh(\omega) - \sinh(\omega)\gamma_0\frac{\vec{k}.\vec{\gamma}}{|\vec{k}|})(\frac{1+\gamma_0}{2})$$

$$\left[(\cosh(\omega) + \sinh(\omega)\gamma_0\frac{\vec{k}.\vec{\gamma}}{|\vec{k}|})(\vec{\gamma}.\vec{\zeta})\gamma_5 + 2\sinh(\omega)\gamma_0\gamma_5\frac{(\vec{k}.\vec{\zeta})}{|\vec{k}|}\right]$$

$$= (\cosh(\omega) - \sinh(\omega)\gamma_0\frac{\vec{k}.\vec{\gamma}}{|\vec{k}|})(\frac{1+\gamma_0}{2})(\cosh(\omega) + \sinh(\omega)\gamma_0\frac{\vec{k}.\vec{\gamma}}{|\vec{k}|})(\vec{\gamma}.\vec{\zeta})\gamma_5$$

$$+ 2\sinh(\omega)(\cosh(\omega) - \sinh(\omega)\gamma_0\frac{\vec{k}.\vec{\gamma}}{|\vec{k}|})(\frac{1+\gamma_0}{2})\gamma_0\gamma_5\frac{(\vec{k}.\vec{\zeta})}{|\vec{k}|}$$

$$= \frac{\slashed{k}+m}{2m}(\vec{\gamma}.\vec{\zeta})\gamma_5 + 2\sinh(\omega)(\cosh(\omega) - \sinh(\omega)\gamma_0\frac{\vec{k}.\vec{\gamma}}{|\vec{k}|})(\frac{1+\gamma_0}{2})\gamma_0\gamma_5\frac{(\vec{k}.\vec{\zeta})}{|\vec{k}|}$$

(A.34)

The second term in the last line in (A.34) can be further simplified by inserting the identity in the form of the product of two exponentials:

$$2\sinh(\omega)(\cosh(\omega)-\sinh(\omega)\gamma_0\frac{\vec{k}.\vec{\gamma}}{|\vec{k}|})(\frac{1+\gamma_0}{2})\gamma_0\gamma_5\frac{\vec{k}.\vec{\zeta}}{|\vec{k}|}$$

$$=2\sinh(\omega)(\cosh(\omega)-\sinh(\omega)\gamma_0\frac{\vec{k}.\vec{\gamma}}{|\vec{k}|})(\frac{1+\gamma_0}{2})(\cosh(\omega)$$

$$+\sinh(\omega)\gamma_0\frac{\vec{k}.\vec{\gamma}}{|\vec{k}|})(\cosh(\omega)-\sinh(\omega)\gamma_0\frac{\vec{k}.\vec{\gamma}}{|\vec{k}|})\gamma_0\gamma_5\zeta_{\parallel}$$

$$=2(\frac{\not{k}+m}{2m})(\frac{\sinh(\omega)\cosh(\omega)}{|\vec{k}|}-\sinh(\omega)^2\gamma_0\frac{\vec{k}.\vec{\gamma}}{|\vec{k}|^2})\gamma_0\gamma_5|\vec{k}|\zeta_{\parallel}$$

$$=2(\frac{\not{k}+m}{2m})(\frac{\sinh(\omega)\cosh(\omega)}{|\vec{k}|}-\sinh(\omega)^2\gamma_0\frac{|\vec{k}|\gamma_{\parallel}}{|\vec{k}|^2})\gamma_0\gamma_5|\vec{k}|\zeta_{\parallel}$$

$$=-2(\frac{\not{k}+m}{2m})\frac{1}{2m}(1+\gamma_0\frac{|\vec{k}|\gamma_{\parallel}}{m+k_0})\gamma_0\gamma_5|\vec{k}|\zeta_{\parallel}$$

$$=-2(\frac{\not{k}+m}{2m})\frac{1}{2m}(1+\gamma_0\frac{|\vec{k}|\gamma_{\parallel}}{m+k_0})\gamma_0\gamma_5|\vec{k}|\zeta_{\parallel} \quad (A.35)$$

Finally we insert(A.35) into(A.34) and get

$$\frac{\not{k}+m}{2m}(\vec{\gamma}.\vec{\zeta})\gamma_5-2(\frac{\not{k}+m}{2m})\frac{1}{2m}(1+\gamma_0\frac{|\vec{k}|\gamma_{\parallel}}{m+k_0})\gamma_0\gamma_5|\vec{k}|\zeta_{\parallel}$$

$$=\frac{\not{k}+m}{2m}\left[\vec{\gamma}_{\perp}.\vec{s}_{\perp}+\gamma_{\parallel}\zeta_{\parallel}-\frac{1}{m}(|\vec{k}|\gamma_0\zeta_{\parallel}-\frac{|\vec{k}|^2\gamma_{\parallel}\zeta_{\parallel}}{m+k_0})\right]\gamma_5$$

$$=\frac{\not{k}+m}{2m}\left[\vec{\gamma}_{\perp}.\vec{s}_{\perp}+\gamma_{\parallel}\zeta_{\parallel}-\frac{1}{m}|\vec{k}|\gamma_0\zeta_{\parallel}+(-1+\frac{k_0}{m})\gamma_{\parallel}\zeta_{\parallel})\right]\gamma_5 \quad (A.36)$$

$$=\frac{\not{k}+m}{2m}\left[\vec{\gamma}_{\perp}.\vec{s}_{\perp}+\gamma_{\parallel}\frac{k_0}{m}\zeta_{\parallel}-\gamma_0\frac{|\vec{k}|}{m}\zeta_{\parallel}\right]\gamma_5$$

$$=\frac{\not{k}+m}{2m}\gamma_5\not{s}$$

Putting (A.36)and(A.31) together we get the desired result(A.30).

**Appendix B)** We use a **Maple** Package[7] for the calculation of the trace of Dirac gamma matrices, to write down the following program.

```
> #Computaion of the current u(bar)(k',s')gamma(3)u(k,s)
> restart;
spur_init():
unprotect(D);
assign(p2[0]=p1[0],k1[0]=0,k2[0]=0,k1[3]=p1[3],k1[1]=p1[1],k1[
2]=p1[2],k2[1]=p2[1],k2[2]=p2[2],k2[3]=p2[3],A=(m+p1[0])/(2*m)
,B=-1/(2*m),D=1/(2*m*(m+p1[0])),p1[1]=0,p1[2]=0);
vectors(p1,p2,s1,s2,k1,k2,t2):
>
definemore(sc4,sc4(p1,p1)=m^2,sc4(p2,p2)=m^2,sc4(p1,s1)=0,sc4(
p2,s2)=0,sc4(s2,s2)=-1,sc4(s1,s1)=-
1,sc4(p1,s1)=0,sc4(p2,k2)=sc4(k2,k2),sc4(k1,k1)=sc4(k2,k2),sc4
(p2,k1)=sc4(p1,k2),sc4(k2,k1)=sc4(k2,p1),sc4(p1,k1)=sc4(k1,k1)
,sc4(k2,k2)=-p1[0]^2+m^2,sc4(k1,s2)=-p1s2,sc4(k2,s2)=-
p2s2,sc4(p1,p2)=p1[0]^2-p1p2,sc4(p1,k2)=-p1p2);
g:=gamma_:
R:=A+B*g[0]&*Z(k1)-B*g[0]&*Z(k2)+D*Z(k1)&*Z(k2):
f:=&*((Z(p1)+m)/(2*m),R,(Z(p2)+m)/(2*m),(1+g[5]&*Z(s2))/2,g[mu
]):
SP(f):
subs(mu=3,%):
simplify(%):
collect(%,[p1[3],p2[3],p1[0]]):
simplify(subs(p1[0]^2=m^2+p1[3]^2,%));
```

$$-\frac{1}{2}\frac{-p1_3-p2_3-I\,p2_2\,s2_1+I\,p2_1\,s2_2}{m}$$